\documentclass[11pt]{article}
\usepackage{epsf,amsmath,amssymb,axodraw,graphicx,scalefnt}
\usepackage{dsfont}
\usepackage{color}
\usepackage{rotating}
\newcommand{\bare}{{\rm B}}
\newcommand{\epscalar}{$\epsilon$-scalar}

\newcommand{\qcd}{{\abbrev QCD}}

\newcommand{\ddim}{\hat}
\newcommand{\edim}{\bar}
\newcommand{\abbrev}{\scalefont{.9}}
\newcommand{\ep}{\epsilon}

\newcommand{\eqn}[1]{Eq.\,(\ref{#1})}
\newcommand{\fig}[1]{Fig.\,\ref{#1}}

\newcommand{\sct}[1]{Sect.\,\ref{#1}}
\newcommand{\dd}{{\rm d}}

\newcommand{\order}[1]{{\cal O}(#1)}

\newcommand{\mssm}{{\abbrev MSSM}}
\newcommand{\susy}{{\abbrev SUSY}}
\newcommand{\bsm}{{\abbrev BSM}}

\newcommand{\dred}{{\abbrev DRED}}
\newcommand{\dreg}{{\abbrev DREG}}
\newcommand{\msbar}{\overline{\mbox{\abbrev MS}}}
\newcommand{\drbar}{\overline{\mbox{\abbrev DR}}}

\title{
  \vskip-3cm{\baselineskip14pt
    \begin{flushleft}
      \normalsize SFB/CPP-09-36 \\
      \normalsize TTP09-13 \\
      \normalsize WUB/09-02 \\
    \end{flushleft}}
  \vskip1.5cm
The \susy{}-\qcd{} $\beta$ function to three loops}
\author{Robert V. Harlander$^{(1)}$, 
  Luminita Mihaila$^{(2)}$,
  Matthias Steinhauser$^{(2)}$\\[2em]
  {\it (1) Fachbereich C, Bergische Universit\"at Wuppertal,}\\
  {\it D-42097 Wuppertal, Germany}\\
  {\tt robert.harlander@uni-wuppertal.de}\\
  {\it (2) Institut f\"ur Theoretische Teilchenphysik, Universit\"at
    Karlsruhe,}\\
  {\it Karlsruhe Institute of Technology (KIT), D-76128 Karlsruhe, Germany}\\
  {\tt luminita@particle.uni-karlsruhe.de, matthias.steinhauser@uka.de}\\
  }

\date{}

\begin{document}
\maketitle


\begin{abstract}
A number of $\drbar$ renormalization constants in softly broken
\susy{}-\qcd{} are evaluated to three-loop level: the wave function
renormalization constants for quarks, squarks, gluons, gluinos, ghosts, 
and \epscalar{}s, and the renormalization constants for the quark and
  gluino mass as well as for all cubic vertices. The latter allow us to
derive the corresponding $\beta$ functions through three loops, all of
which we find to be identical to the expression for the gauge $\beta$
function obtained by Jack, Jones, and North~\cite{Jack:1996vg} (see also
Ref.\,\cite{Pickering:2001aq}). This explicitly demonstrates the
consistency of \dred{} with \susy{} and gauge invariance, an important
pre-requisite for precision calculations in supersymmetric theories.
\end{abstract}



\section{Introduction}

Up to now, particle collider experiments could not provide clear
evidence for physics beyond the Standard Model (\bsm{}). The quest for
extensions of the Standard Model comes either from non-accelerator
observations, such as Dark Matter and Dark Energy, or it is based on
purely theoretical considerations. For example,
naturalness~\cite{Susskind:1978ms, Weinberg:1979bn, tHooft:1979bh}
requires some mechanism to stabilize 
the Higgs mass at the electro-weak scale (fine tuning problem); the
intriguing evolution of the gauge couplings towards a common value at
high energies needs an enlarged particle spectrum if unification is to
occur at scales compatible with proton
decay~\cite{Ellis:1990wk,Amaldi:1991cn,Langacker:1991an}; any explanation for
electro-weak symmetry breaking presumably requires an embedding of the
Standard Model into a higher symmetry.

In the light of these considerations, supersymmetry (\susy{}) is a
strong candidate for an extension of the Standard
Model~\cite{Nilles:1983ge}. Not only does it provide solutions for Dark
Matter, it also solves the fine tuning problem by canceling the
quadratic divergences in the Higgs self energies, it explains
electro-weak symmetry breaking by a simple evolution of the parameters
of the Higgs potential, and unification of the gauge couplings can be
achieved for rather natural choices of the parameters.

This latter issue is a rather unique feature: it actually allows one to
make quantitative predictions at energy scales that are several orders
of magnitude larger than what can be achieved at current and probably
also at any future particle collider. Crucial ingredients for such
indirect measurements are precision data as well as precision
calculations (see, e.g., Ref.\,\cite{Harlander:2007wh}).

In order to study the unification of the gauge couplings in \susy{}, one
needs to include the effect of the supersymmetric particle spectrum into
the running. In addition, dimensional regularization with minimal
subtraction, called $\msbar$, is no longer a good renormalization scheme
in the sense that it explicitly breaks supersymmetry. An alternative
regularization, so-called Dimensional Reduction (\dred{}) was suggested
by Siegel~\cite{Siegel:1979wq}, however, inconsistencies of this method
were pointed out only shortly
afterwards~\cite{Siegel:1980qs}. Nevertheless, renormalization by
combining \dred{} with minimal subtraction (the $\drbar{}$ scheme) has
become the preferred scheme in higher order supersymmetric
calculations~\cite{Aguilar-Saavedra:2005pw}. A thorough formulation of
\dred{} which isolates the source of the inconsistencies has been given
in Ref.\,\cite{Stockinger:2005gx}. It was pointed out that, although a
mathematical consistent formulation of \dred{} is possible, it could
violate \susy{} at higher orders of
perturbation theory. It is the main goal of the current paper to
demonstrate this consistency within \susy{}-\qcd{} up to the three-loop
level.

Renormalization group functions, governing the energy dependence of
masses and couplings, are among the simplest quantities to compute in
perturbative quantum field theory. For example, the anomalous dimensions
of the strong coupling constant in \qcd{} (the \qcd{} $\beta$ function)
and the quark masses are known at four-loop level in the $\msbar$
scheme~\cite{vanRitbergen:1997va,Chetyrkin:1997dh,Vermaseren:1997fq,Czakon:2004bu}
(the conversion to the $\drbar$ scheme was done in
Ref.\,\cite{Harlander:2006xq}).


For supersymmetric gauge theories, one can devise a particular
renormalization scheme, so-called {\abbrev NSVZ}~\cite{Novikov:1983uc},
where an all-order relation between the gauge $\beta$ function and the
anomalous dimension of the chiral supermultiplet is valid. So, in the
absence of the matter supermultiplet, {\it i.e.} for a \susy{} Yang-Mills
theory, the $\beta$ function is known to all orders in the coupling
constant.  The problem with this scheme is that general conversion rules
to schemes used in perturbative calculations such as $\msbar$ or
$\drbar$ are not known.  However, by explicit calculation of the abelian
$\beta$ function in the $\drbar$ scheme, comparison with the {\abbrev
  NSVZ} result, and arguments based on its holomorphy, the authors of
Ref.\,\cite{Jack:1996vg} found the two-loop conversion formula for the
coupling also in the non-abelian case. The \susy{}-\qcd{} $\beta$
function in the $\drbar$ scheme is thus known to three loops. The result
was later confirmed by an explicit calculation using the background
field method~\cite{Pickering:2001aq}, which requires three-loop
corrections to the gluon propagator.

Applying the same method based on holomorphy of the {\abbrev NSVZ}
scheme to softly broken \susy{} gauge theory, the authors of
Ref.\,\cite{Hisano:1997ua} derived the renormalization group equation
governing the running of the gaugino and squark masses, valid
to all orders in the coupling constant. Using this result, an elegant
relation between the gaugino and the gauge $\beta$ functions was
formulated in Ref.\,\cite{Jack:1997pa}, both for the {\abbrev NSVZ} and the
$\drbar$ scheme.

One of the goals of the current paper is to add another confirmation of
these results by calculating for the first time directly the relevant
vertex diagrams through three-loop order. Dimensional reduction is
implemented by explicitly introducing \epscalar{}s, the
$(4-d)$-dimensional components of the gluon field.

In addition, we derive the gauge $\beta$ function $\beta_s$ from all
possible three-point functions in \susy{}-\qcd{} that involve the strong
coupling $\alpha_s$. The fact that in each case we obtain the same
expression provides an explicit check on the consistency of \dred{} with
gauge invariance and supersymmetry. In particular, we calculate the
renormalization of the quark-squark-gluino vertex which is not fixed by
gauge invariance but by supersymmetry alone.
Furthermore, we confirm the relation between  $\beta_s$ and the gluino
mass anomalous dimension $\gamma_{\tilde g}$ by
an explicit calculation of $\gamma_{\tilde g}$ to three loops.
We also verify the three-loop formula for the quark anomalous dimension
$\gamma_{q}$ ~\cite{Jack:1996qq} that was derived in the superfield formalism.  
It is 
an essential 
ingredient for the derivation of the four-loop \susy{}-\qcd{}
gauge $\beta$ function in the approach of Ref.~\cite{Jack:1996qq}.

The \epscalar{}s occurring in \dred{} have Feynman rules that can be
derived from 
the ones of the gluon by replacing $d$-dimensional objects with
$2\ep$-dimensional ones. In a supersymmetric theory, the coupling
constants for the \epscalar{} and the gluon must be identical, which
means that the $\beta$ function for the \epscalar{} couplings must be
given by the one of the strong coupling constant $\alpha_s$. We check
also this relation through three loops, by evaluating three-point
functions involving \epscalar{}s through three loops.

The remainder of this paper is organized as follows: In
\sct{sec::dred}, we review the method of \dred{} and describe its
implementation in our setup.  \sct{sec::method} describes some details
of the calculation for the Green's functions and the $\beta$ functions,
and also gives the main results.  Concluding remarks are given in
\sct{sec::conclusions}.


\section{Dimensional Reduction}\label{sec::dred}

\subsection{Notation and technical setup}

We implement \dred{} by introducing the quasi-four, -$d$, and
-$2\ep$-dimensional spaces~\cite{Avdeev:1981vf}  (see also 
  Ref.\,\cite{Stockinger:2005gx}) $Q_{4}$,
$Q_{d}$, $Q_{2\ep}$, where $d=4-2\ep$ and
\begin{equation}
\begin{split}
Q_{4} = Q_{d} \cup Q_{2\ep}\qquad \mbox{and }\quad
Q_{d} \cap Q_{2\ep} = 0\,.
\end{split}
\end{equation}
Also, we introduce quasi-$four$, -$d$ and -$2\ep$-dimensional sets of indices,
implying the correspondence 
\begin{equation}
\begin{split}
\mu\leftrightarrow (\ddim\mu,\edim\mu)\,,\quad
\nu\leftrightarrow (\ddim\nu,\edim\nu)\,,\quad
\rho\leftrightarrow (\ddim\rho,\edim\rho)\,,\quad
\sigma\leftrightarrow (\ddim\sigma,\edim\sigma)\,,
\end{split}
\end{equation}
in the sense that, for example,
\begin{equation}
\begin{split}
t^{\mu} = t^{\ddim\mu} + t^{\edim\mu}\qquad \mbox{etc.}
\end{split}
\end{equation}
for any vector $t$. I.e., $t^{\ddim\mu}$ and $t^{\edim\mu}$ denote the projections
of $t^{\mu}\in Q_4$ to the subspaces $Q_{d}$ and $Q_{2\ep}$, respectively.
For generic indices without reference to (quasi-)dimensionality we will
use the letters $\alpha,\beta,\gamma,\delta$.

For the metric tensor, we have
\begin{equation}
\begin{split}
g^{\ddim\mu\edim\mu} &= 0\qquad\Rightarrow \qquad
g^{\mu\nu} = g^{\ddim\mu\ddim\nu} + g^{\edim\mu\edim\nu}\,,\\
g^{\edim\mu\edim\nu}t_{\mu} &= t^{\edim\nu}\,,\qquad
g^{\ddim\mu\ddim\nu}t_{\mu} = t^{\ddim\nu}\,,\qquad
g^{\mu}_{\mu} = 4\,,\qquad
g^{\ddim\mu}_{\ddim\mu} = d\,,\qquad
g^{\edim\mu}_{\edim\mu} = 2\ep\,.
\end{split}
\end{equation}

The usual commutation relations for four-dimensional Dirac matrices are
carried forward to their $d$- and $2\ep$-dimensional projections:
\begin{equation}
\begin{split}
\{\gamma^{\ddim\mu},\gamma^{\ddim\nu}\} = 2g^{\ddim\mu\ddim\nu}\,,\qquad
\{\gamma^{\edim\mu},\gamma^{\edim\nu}\} = 2g^{\edim\mu\edim\nu}\,,\qquad
\{\gamma^{\ddim\mu},\gamma^{\edim\nu}\} = 0\,.
\label{eq::antigamma}
\end{split}
\end{equation}
These relations together with the constraint ${\rm
  Tr}\, \mathds{1} =4 $ allow us to evaluate fermion 
traces (without $\gamma_5$) in the following way: 
\begin{equation}
\begin{split}
&{\rm
  Tr}(\gamma^{\ddim\mu_1}\cdots\gamma^{\ddim\mu_n}\,
  \gamma^{\edim\nu_1}\cdots\gamma^{\edim\nu_m}) 
= \frac{1}{4}
{\rm Tr}(\gamma^{\ddim\mu_1}\cdots\gamma^{\ddim\mu_n})
{\rm Tr}(\gamma^{\edim\nu_1}\cdots\gamma^{\edim\nu_m})\,.
\label{eq::traces}
\end{split}
\end{equation}
However, the quark-squark-gluino vertices involve $\gamma_5$, and the
problems involving $\gamma_5$ in other than four dimensions are of
course well known. We treat it as follows:
\begin{enumerate}
\item Use the relations
  \begin{equation}
\begin{split}
  \{\gamma^{\ddim\mu},\gamma_5\} = \{\gamma^{\edim\mu},\gamma_5\} = 0\,,\qquad
  (\gamma_5)^2 = 1\,,
\label{eq::trgamma5}
\end{split}\end{equation}
  until there is at most one $\gamma_5$ per fermion trace.
  If all $\gamma_5$ matrices are eliminated in this way, one can
  continue with \eqn{eq::traces}.
\item Fermion traces with a
  single $\gamma_5$ matrix are reduced to the form 
  ${\rm Tr}(\gamma^\alpha\gamma^\beta\gamma^\gamma\gamma^\delta\gamma_5)$
  by using Eqs.~(\ref{eq::antigamma}) and (\ref{eq::trgamma5}), 
  and then evaluated by applying the 
  {\it formal} replacement
\begin{equation}
\begin{split}
  {\rm Tr}(\gamma^\alpha\gamma^\beta\gamma^\gamma\gamma^\delta\gamma_5)
  = 4i\,\varepsilon^{\alpha\beta\gamma\delta}\,.
  \label{eq::trgamma5_2}
\end{split}\end{equation}
\item The intrinsically ({\it not} quasi-)four-dimensional Levi-Civita
  symbol $\varepsilon^{\alpha\beta\gamma\delta}$ in \eqn{eq::trgamma5_2}
  with quasi-dimensional indices makes sense when combined with a second
  one of its kind. By applying appropriate projectors, our calculations
  are set up in such a way that this is always the case, so we can use
  \begin{equation}
\begin{split}
  \varepsilon^{\alpha\beta\gamma\delta}
  \varepsilon_{\alpha'\beta'\gamma'\delta'} = 
  g^{[\alpha\phantom{]}}_{[\alpha'\phantom{]}}
  g^{\phantom{[}\beta\phantom{]}}_{\phantom{[}\beta'\phantom{]}} 
  g^{\phantom{[}\gamma\phantom{]}}_{\phantom{[}\gamma'\phantom{]}} 
  g^{\phantom{[}\delta\,]}_{\phantom{[}\delta']}\,,
\end{split}\end{equation}
    where the square brackets denote anti-symmetrization. Note that
    due to $g^{\ddim\mu\edim\mu}=0$, one can write, e.g.,
    \[
    \begin{split}
    \varepsilon^{\ddim\mu\ddim\nu\ddim\rho\edim\sigma}
    \varepsilon_{\ddim\mu'\ddim\nu'\ddim\rho'\edim\sigma'} =
  g^{[\ddim\mu\phantom{]}}_{[\ddim\mu'\phantom{]}}
  g^{\phantom{[}\ddim\nu \phantom{]}}_{\phantom{[}\ddim\nu'\phantom{]}}
  g^{\phantom{[}\ddim\rho\,]}_{\phantom{[}\ddim\rho']} \,
  g^{\edim\sigma}_{\edim\sigma'}\,,\qquad&
    \varepsilon^{\ddim\mu\ddim\nu\edim\rho\edim\sigma}
    \varepsilon_{\ddim\mu'\ddim\nu'\edim\rho'\edim\sigma'} =
  g^{[\ddim\mu\phantom{]}}_{[\ddim\mu'\phantom{]}}
  g^{\phantom{[}\ddim\nu ]}_{\phantom{[}\ddim\nu']} \,
  g^{[\edim\rho\phantom{]}}_{[\edim\rho'\phantom{]}} 
  g^{\phantom{[}\edim\sigma\,]}_{\phantom{[}\edim\sigma']}\,, \\
    \varepsilon^{\ddim\mu\ddim\nu\ddim\rho\edim\sigma}
    \varepsilon_{\ddim\mu'\ddim\nu'\edim\sigma'\edim\tau'} = 0\,,\qquad {\rm etc.}
    \end{split}
    \]
\end{enumerate}

We made sure that the terms that need to be treated in this way generate
no higher than simple poles in $\ep$. Therefore, we can be sure that the
above procedure is valid without introducing additional finite counterterms.
Of course, the fact that we find the same $\beta$ function from all vertices 
also supports our treatment of $\gamma_5$.


\subsection{\epscalar{}s and corresponding Feynman rules}

We work in $N=1$ supersymmetric \qcd{} with $n_f$ quark flavors,
corresponding to the strong sector of the \mssm{}. The Feynman rules are
therefore given by the usual ones for \qcd{}, involving quarks $q$,
gluons $g$, and ghosts $c$, plus the Feynman rules for their
supersymmetric partners: the squarks $\tilde q_L$, $\tilde q_R$, and the
gluinos, $\tilde g$. In our implementation, instead of the partners of
the left- and right-handed quarks, we will use the mass eigenstates
\[
\left(\begin{matrix}
  \tilde q_1\\\tilde q_2
\end{matrix}\right) = 
\left(\begin{matrix}
  \cos\theta_q & \sin\theta_q\\
  -\sin\theta_q & \cos\theta_q
\end{matrix}\right)
\left(\begin{matrix}
  \tilde q_L\\\tilde q_R
\end{matrix}\right)\,.
\]
The mixing angle $\theta_q$ depends on the squark masses; however, all
the quantities in this paper will be independent of any masses, and
therefore, they must also be independent of the squark mixing
angle. This serves as a welcome check on our calculations.

The tree-level Feynman rules for \susy{}-\qcd{} in \dreg{} can be found
in the literature; in this paper, we will use those given in
Ref.\,\cite{Harlander:2004tp}.  \dred{} results are obtained by
supplementing these Feynman rules by new ones for the so-called
\epscalar{}s, as will be explained in what follows.

\dred{} is defined by assuming that all fields depend only on the
coordinates $x^{\ddim\mu}$ of a $d$-dimensional subspace of the usual
four-dimensional Minkowski space. I.e., the $2\ep$-dimensional
components $x^{\edim\mu}$ have no physical relevance. In particular,
this means that derivatives $\partial_{\mu}$ and momenta
$p_{\mu}$ can always be replaced by $\partial_{\ddim\mu}$ and $p_{\ddim\mu}$.  The
technical implementation is done by decomposing the quasi-four-dimensional
vector fields into $d$- and $2\ep$-dimensional components:
\begin{equation}
\begin{split}
A_{\mu} = A_{\ddim\mu} + A_{\edim\mu}\,.
\label{eq::epscalar}
\end{split}\end{equation}
For convenience, $A_{\ddim\mu}$ will be simply called the ``gluon field'' in
what follows, while $A_{\mu}$ will be explicitly referred to as
``four-dimensional gluon field''. $A_{\edim\mu}$ is the so-called
\epscalar. The corresponding Feynman rules for the \epscalar{}s are
obtained from the $d$-dimensional \susy{}-\qcd{} ones by replacing
the $d$-dimensional indices by $2\ep$-dimensional ones. In this way, one
arrives at the following vertices: $q\bar q\ep$, $g\ep\ep$, $\tilde
g\tilde g\ep$, $\ep\ep\tilde q\tilde q$, $gg\ep\ep$, $\ep\ep\ep\ep$,
where $\ep$ denotes the \epscalar{}.  Their Feynman rules are
given explicitly in Appendix~\ref{app::frules}.

The bare coupling constants for the gluon and the \epscalar{} are
identical by construction. However, {\it a priori} it is not clear
whether this holds also for the renormalized couplings. In principle,
all \epscalar{} couplings could be different without violating gauge
invariance.  In fact, it is well known that they differ in standard
(i.e., non-\susy{}) \qcd{} (see, e.g.,
Refs.\,\cite{Harlander:2006rj,Harlander:2006xq,Jack:2007ni}). Even more:
in order to renormalize the quartic \epscalar{} vertex, one has to take
into account all possible color structures for it, and attribute to
each one a separate coupling constant. For \susy{}-\qcd{}, we have
explicitly checked that at the one-loop order only the $\beta$ function
associated with the usual color structure of the four-gluon interaction,
{\it i.e.} $f_{abe} f_{cde}$\footnote{$f_{abe}$ denotes the structure
  constants of the gauge group.}, does not vanish and it equals the
one-loop gauge $\beta$ function.  So, through one-loop, one can identify
the coupling constant of the corresponding \epscalar{} quartic
interaction with the strong coupling constant and set to zero the other
three couplings. This order of accuracy is sufficient for the results
discussed in this paper, as the \epscalar{} quartic interactions
contribute to the anomalous dimensions starting from the two-loop order.

In order for the renormalized Lagrangian to obey \susy{}, the
decomposition of \eqn{eq::epscalar} should hold also at higher orders.
Therefore, the renormalized gluon and \epscalar{} coupling constants
must be equal, i.e., their $\beta$ functions must be the same.  We will
explicitly verify this for the $q\bar q\ep$, the $\tilde g\tilde
g\ep$, and the $g\ep\ep$ vertex through three loops, thus confirming
consistency of \dred{} with \susy{} at 
  next-to-next-to-next-to-leading order of perturbation theory.

\subsection{Majorana fermions}\label{sec::majorana}

Another technical issue is given by the Majorana character of the
gluinos. A well-defined and practical prescription how to deal with
Majoranas in higher order perturbative calculations was given  in
  Ref.~\cite{Denner:1992me,Denner:1992vza}. Unfortunately, using this
method in combination with the Feynman diagram generator {\tt
  qgraf}~\cite{Nogueira:1993ex} is not straightforward. We implement it
by parsing the output of {\tt qgraf} with a {\tt PERL} program, giving
each chain of fermion lines a well-defined direction. If the chain
contains a Dirac fermion $f$ whose ordinary fermion direction (``charge
flow'') is opposite to the one chosen by the {\tt PERL} program
  (``external flow''), we replace this particle with a ``flipped
fermion'' $\check f$ which has opposite charge flow. Majoranas have no
definite charge flow and are simply interpreted as Dirac particles with
charge flow equal to the external flow.


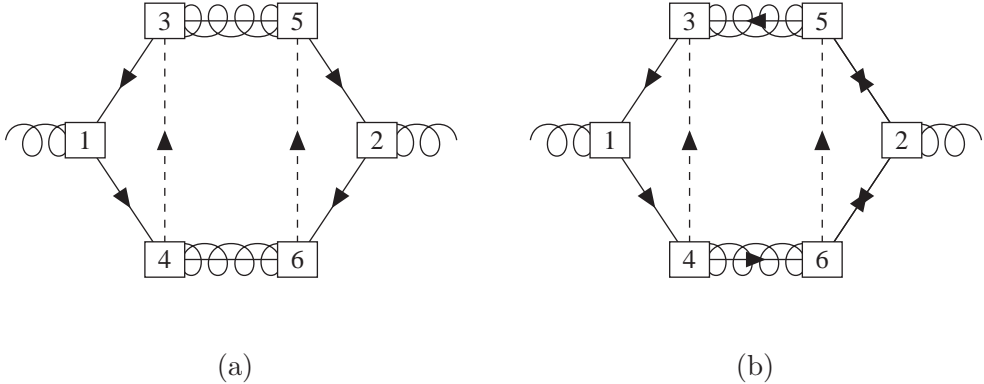
\begin{figure}
\begin{center}
\newcommand{\qvertex}[2]{%
\ifnum#1=-1 \ifnum#2=1 0 \else\ifnum#2=2 0\fi\fi\else
\ifnum#1=6
    \ifnum#2=1 100
\else\ifnum#2=2 30
\fi\fi\else
\ifnum#1=4
    \ifnum#2=1 50
\else\ifnum#2=2 30
\fi\fi\else
\ifnum#1=1
    \ifnum#2=1 20
\else\ifnum#2=2 75
\fi\fi\else
\ifnum#1=3
    \ifnum#2=1 50
\else\ifnum#2=2 120
\fi\fi\else
\ifnum#1=2
    \ifnum#2=1 130
\else\ifnum#2=2 75
\fi\fi\else
\ifnum#1=5
    \ifnum#2=1 100
\else\ifnum#2=2 120
\fi\fi\else
\fi\fi\fi\fi\fi\fi\fi }
\begin{tabular}{cc}
\begin{picture}(150,150)
\SetWidth{0.5}
\ArrowLine(\qvertex{3}{1},\qvertex{3}{2})(\qvertex{1}{1},\qvertex{1}{2})
\SetWidth{0.5}
\ArrowLine(\qvertex{1}{1},\qvertex{1}{2})(\qvertex{4}{1},\qvertex{4}{2})
\SetWidth{0.5}
\ArrowLine(\qvertex{5}{1},\qvertex{5}{2})(\qvertex{2}{1},\qvertex{2}{2})
\SetWidth{0.5}
\ArrowLine(\qvertex{2}{1},\qvertex{2}{2})(\qvertex{6}{1},\qvertex{6}{2})
\SetWidth{0.5}
\DashArrowLine(\qvertex{4}{1},\qvertex{4}{2})(\qvertex{3}{1},\qvertex{3}{2}){3}
\SetWidth{0.5}
\DashArrowLine(\qvertex{6}{1},\qvertex{6}{2})(\qvertex{5}{1},\qvertex{5}{2}){3}
\SetWidth{0.5}
\Line(\qvertex{5}{1},\qvertex{5}{2})(\qvertex{3}{1},\qvertex{3}{2})
\SetWidth{0.5}
\Gluon(\qvertex{5}{1},\qvertex{5}{2})(\qvertex{3}{1},\qvertex{3}{2}){6}{4}
\SetWidth{0.5}
\Line(\qvertex{4}{1},\qvertex{4}{2})(\qvertex{6}{1},\qvertex{6}{2})
\SetWidth{0.5}
\Gluon(\qvertex{4}{1},\qvertex{4}{2})(\qvertex{6}{1},\qvertex{6}{2}){6}{4}
\SetWidth{0.5}
\Gluon(-10,75)(\qvertex{1}{1},\qvertex{1}{2}){6}{2}
\SetWidth{0.5}
\Gluon(\qvertex{2}{1},\qvertex{2}{2})(160,75){6}{2}
\CText(\qvertex{6}{1},\qvertex{6}{2}){Black}{White}{6}
\CText(\qvertex{4}{1},\qvertex{4}{2}){Black}{White}{4}
\CText(\qvertex{1}{1},\qvertex{1}{2}){Black}{White}{1}
\CText(\qvertex{3}{1},\qvertex{3}{2}){Black}{White}{3}
\CText(\qvertex{2}{1},\qvertex{2}{2}){Black}{White}{2}
\CText(\qvertex{5}{1},\qvertex{5}{2}){Black}{White}{5}
\end{picture} 
\hspace*{3em}&
\begin{picture}(150,150)
\SetWidth{0.5}
\ArrowLine(\qvertex{3}{1},\qvertex{3}{2})(\qvertex{1}{1},\qvertex{1}{2})
\SetWidth{0.5}
\ArrowLine(\qvertex{1}{1},\qvertex{1}{2})(\qvertex{4}{1},\qvertex{4}{2})
\SetWidth{0.5}
\ArrowLine(\qvertex{2}{1},\qvertex{2}{2})(\qvertex{5}{1},\qvertex{5}{2})
\SetWidth{0.5}
\ArrowLine(\qvertex{5}{1},\qvertex{5}{2})(\qvertex{2}{1},\qvertex{2}{2})
\SetWidth{0.5}
\ArrowLine(\qvertex{6}{1},\qvertex{6}{2})(\qvertex{2}{1},\qvertex{2}{2})
\SetWidth{0.5}
\ArrowLine(\qvertex{2}{1},\qvertex{2}{2})(\qvertex{6}{1},\qvertex{6}{2})
\SetWidth{0.5}
\DashArrowLine(\qvertex{4}{1},\qvertex{4}{2})(\qvertex{3}{1},\qvertex{3}{2}){3}
\SetWidth{0.5}
\DashArrowLine(\qvertex{6}{1},\qvertex{6}{2})(\qvertex{5}{1},\qvertex{5}{2}){3}
\SetWidth{0.5}
\ArrowLine(\qvertex{5}{1},\qvertex{5}{2})(\qvertex{3}{1},\qvertex{3}{2})
\SetWidth{0.5}
\Gluon(\qvertex{5}{1},\qvertex{5}{2})(\qvertex{3}{1},\qvertex{3}{2}){6}{4}
\SetWidth{0.5}
\ArrowLine(\qvertex{4}{1},\qvertex{4}{2})(\qvertex{6}{1},\qvertex{6}{2})
\SetWidth{0.5}
\Gluon(\qvertex{4}{1},\qvertex{4}{2})(\qvertex{6}{1},\qvertex{6}{2}){6}{4}
\SetWidth{0.5}
\Gluon(-10,75)(\qvertex{1}{1},\qvertex{1}{2}){6}{2}
\SetWidth{0.5}
\Gluon(\qvertex{2}{1},\qvertex{2}{2})(160,75){6}{2}
\CText(\qvertex{6}{1},\qvertex{6}{2}){Black}{White}{6}
\CText(\qvertex{4}{1},\qvertex{4}{2}){Black}{White}{4}
\CText(\qvertex{1}{1},\qvertex{1}{2}){Black}{White}{1}
\CText(\qvertex{3}{1},\qvertex{3}{2}){Black}{White}{3}
\CText(\qvertex{2}{1},\qvertex{2}{2}){Black}{White}{2}
\CText(\qvertex{5}{1},\qvertex{5}{2}){Black}{White}{5}
\end{picture} 
\\
(a)\hspace*{3em}& (b)
\end{tabular}
\caption[]{\label{fig::majorana}(a) Feynman diagram involving Majorana
  fermions in a non-trivial way. Solid lines are quarks, dashed lines
  are squarks, slashed springy lines are gluinos, and the external lines
  are gluons. The arrows on the lines denote the charge flow. Note that
  the fermion line involves quarks with opposite charge flow.  (b) The
  same diagram after applying the prescription proposed in
    Ref.~\cite{Denner:1992me,Denner:1992vza}.  The arrows on the gluino
  lines denote the external flow (see text); lines with two
  ``clashing'' arrows are ``flipped'' quarks which have  external flow
  opposite to the charge flow.}
\end{center}
\end{figure}

\begin{figure}
\begin{verbatim}
          |1|g                                   |1|g        
          |2|g                                   |2|g        
          |3,1|fQ,fq                             |3,1|fQ,fq  
          |1,4|fQ,fq                             |1,4|fQ,fq  
          |5,2|fQ,fq                             |2,5|fQx,fqx  
          |2,6|fQ,fq                             |6,2|fQx,fqx  
          |4,3|Sq1,sq1                           |4,3|Sq1,sq1
          |5,3|fgm,fgm                           |5,3|fgmx,fgm
          |6,4|fgm,fgm                           |4,6|fgmx,fgm
          |6,5|Sq1,sq1                           |6,5|Sq1,sq1 
\end{verbatim}
\caption[]{\label{fig::majqgraf}The diagrams of \fig{fig::majorana} in
  topological notation.}
\end{figure}


An example is shown in \fig{fig::majorana}. In topological notation,
diagram~(a) is represented by the left column of \fig{fig::majqgraf},
where each line denotes a propagator in obvious notation. After
application of the prescription of
  Ref.~\cite{Denner:1992me,Denner:1992vza}, the diagram in 
\fig{fig::majorana}\,(b) results, or, again in topological notation, the
right column of \fig{fig::majqgraf}. One can see that the quarks ({\tt
  fQ,fq}) connecting vertices 2 with 5 and 6 with 2 have been replaced
by ``flipped'' quarks ({\tt fQx,fqx}), and the Majorana notation for the
gluinos ({\tt fgm,fgm}) has been changed to the Dirac notation, where
particle and anti-particle are distinguishable ({\tt fgmx,fgm}).

The result looks like regular {\tt qgraf} output with only Dirac
fermions. The flipped fermions have Feynman rules that are easily
derived from the original fermions using
Ref.\,\cite{Denner:1992me,Denner:1992vza}.  They are given
  explicitly in Appendix~\ref{app::frules}. Finally, the overall sign
is determined in the usual way by counting the number of closed fermion
loops.

Note that the same approach has already been applied successfully to
earlier calculations~\cite{Harlander:2008ju}.

For the remainder of the calculation, we use the following setup: the
diagrams are brought to {\tt MINCER} notation~\cite{Larin:1991fz} with
the help of the {\tt C++} programs {\tt q2e} and {\tt exp}~\cite{exp}. The actual
calculation is done in {\tt FORM}~\cite{Vermaseren:2000nd}: After
applying appropriate projectors in Lorentz, Dirac and color space and
evaluating the fermion traces, we use {\tt MINCER} to evaluate the
Feynman integrals in terms of a Laurent series in $\ep$.

The representation of the four-gluon vertex through an auxiliary,
non-propagating particle allows us to factorize the color factor from
the rest of the diagram and to evaluate it separately. For that purpose,
we use the program {\tt color.frm}~\cite{vanRitbergen:1998pn} which
expresses the result in terms of color invariants.


\section{Evaluation of the renormalization constants}\label{sec::method}
We closely follow the method of
Refs.\,\cite{Steinhauser:1998cm,Steinhauser:2002rq}, i.e., we calculate
connected, amputated Green's functions and derive the coupling constant
renormalization from Slavnov-Taylor identities. For example, if we
compute the $N$-point Green's function with external fields
$\phi_1,\cdots,\phi_n$ and denote its coupling constant by $g$, we
obtain
\begin{equation}
\begin{split}
Z_g = \frac{Z_{\phi_1\cdots\phi_N}}{\sqrt{Z_{\phi_1}\cdots Z_{\phi_N}}}\,,
\label{eq::ZZZ}
\end{split}\end{equation}
where the $Z_{\phi_i}$ are the wave function renormalization constants
for the $\phi_i$, $Z_{\phi_1\cdots\phi_N}$ is the corresponding vertex
renormalization constant, and $Z_g$ the charge renormalization.

Since we work in a minimal subtraction scheme, the renormalization
constants (and thus also the $\beta$ function) are independent of any
mass scale.  We remind the reader that due to this fact, decoupling of
heavy particles has to be done ``by hand'' in the $\drbar{}$ scheme, by
introduction of decoupling constants for the renormalized parameters of
the model.  The decoupling constants in the \mssm{} for the strong
coupling constant and the quark mass have been calculated to two-loop
order in Refs.\,\cite{Harlander:2007wh,Bauer:2008bj}.

The anomalous dimension for the coupling $g$ can be obtained
from the simple pole of $Z_g$ in the usual way via
\begin{equation}
\begin{split}
0 \equiv \mu\frac{\dd}{\dd\mu} g^\bare =
\mu\frac{\dd}{\dd\mu}\left(\mu^\ep Z_g g\right)\,,
\end{split}
\end{equation}
where $g^\bare$ is the bare coupling, and $\mu$ is the (arbitrary)
renormalization scale. $Z_g$ does not explicitly depend on $\mu$, and
if $g$ is the only coupling constant in the theory, then
\begin{equation}
\begin{split}
\mu\frac{\dd}{\dd\mu} g = \frac{-\ep g}{1+g\frac{\dd}{\dd g}Z_g}\,.
\label{eq::gZg}
\end{split}
\end{equation}
Applying \dred{} to a non-\susy{} theory like standard \qcd{}, $Z_g$ in
general depends also on the \epscalar{} couplings, and \eqn{eq::gZg}
assumes a slightly more complicated form (see, e.g.,
Ref.\,\cite{Harlander:2006rj}).
The $\beta$ function is defined through
\begin{equation}
\begin{split}
\beta(\alpha_s) =\mu^2\frac{\dd}{\dd\mu^2}\frac{\alpha_s(\mu^2)
}{\pi}\,,\quad \mbox{where}\,\quad \alpha_s=\frac{g^2}{4\pi}\,.
\label{eq::betaS}
\end{split}
\end{equation}

In analogy to \eqn{eq::betaS} we define the anomalous dimension for
fermion (quark and gluino) masses
\begin{equation}
\begin{split}
\gamma_{i}=\frac{\mu^2}{m_i}\frac{\dd m_i}{\dd\mu^2}  = -\pi \beta_s\frac{\dd
  \ln Z_{m,i}}{\dd \alpha_s}\,,\quad i=q\,,\tilde{g}\,,
\label{eq::gamM}
\end{split}
\end{equation}
where $Z_{m,i}$ is the mass renormalization constant.

Since we work in a \susy{} theory, we expect the renormalization
constants for the \epscalar{} couplings to be identical to the ones of
the strong coupling $g_s$,  provided \dred{} is consistent with
  \susy{}.  When evaluating the strong coupling renormalization $Z_s$
at $n$-loop, at most the $(n-1)$-loop expression for the \epscalar{}
coupling renormalization $Z_e$ is needed. Therefore, the calculation of
$Z_s$ at $n$-loop order checks the identity $Z_s=Z_e$ through
$(n-1)$-loop order: if $Z_s\neq Z_e$ at $(n-1)$-loop order, in most
  cases the $n$-loop expression for $Z_s$ would not be local.  However,
  in some cases it can also happen that simply the $1/\epsilon$
  pole is wrong.

In our calculation it is $n=3$, so we check the equality to two-loop
order.  In addition to that, however, we have explicitly evaluated
 a number of \epscalar{} couplings to three-loop order 
  (see Table~\ref{tab::dianums}) and also find equality to the strong
coupling constant.  In this way, consistency of \dred{} with \susy{} is
confirmed through three-loop order.

\begin{table}
\begin{center}
\begin{tabular}[t]{|c||c|c|c|}
\hline
\# loops & 1 & 2 & 3 \\
\hline
$c\bar c$ &  1 & 14 & 423 \\
$q\bar q$ &  4 & 86 & 3583 \\
$\ep\ep$ &  7 & 100 & 3902 \\
$\tilde g\tilde g$  & 6 & 130 & 5577\\
$\tilde q\tilde q$ &  8 & 157 & 6760\\
$gg$ &  11 & 171 & 6954\\
\hline
\end{tabular}
\hspace*{3em}
\begin{tabular}[t]{|c||c|c|c|c|}
\hline
\# loops &0 & 1 & 2 & 3 \\
\hline
$c\bar c g$ & 1 & 2 & 77 & 3920 \\
$q\bar q\ep$ & 1 & 6 & 319 & 21669 \\
$q\bar qg $ & 1 & 8 & 439 & 30078 \\
$\tilde g\tilde g\ep$ & 1 & 8 & 445 & 31815\\
$q\tilde q\tilde g$ & 1 & 6 & 430 & 36868\\
$\ep\ep g$ & 1 & 14 & 618 & 38741 \\
$\tilde g\tilde g g$ & 1 & 12 & 657 & 46314\\
$\tilde q\tilde q g$ & 1 & 12 & 674 & 52205 \\
$ggg$ & 1 & 26 & 1105 & 70705\\
\hline
\end{tabular}
\end{center}
\caption[]{\label{tab::dianums}The number of diagrams contributing to
  the Green's functions evaluated in this work.  Left table: two-point
  functions; right table: three-point functions. The first column
  indicates the external legs of the Green's function, the other columns
  show the number of diagrams at the individual loop orders. 
  The number depends on the implementation of the
    quartic vertices. In our calculation, we split the $gggg$,
    $gg\ep\ep$, and the $\ep\ep\ep\ep$ each into two cubic vertices by
    introducing non-propagating auxiliary particles.}
\end{table}

For the calculation of the renormalization constants within  the
minimal subtraction scheme, one is free to choose any masses and
external momenta, as long as infra-red divergences are avoided.  We set
all masses to zero, as well as one of the  two independent
external momenta in the three-point functions. We therefore arrive at
three-loop integrals with one non-vanishing external momentum $q$ which
can be calculated with the help of {\tt MINCER}. As a check, we have
also calculated some of the three-point functions  with
  non-zero particle masses $m$ by expanding them in the limit
$m^2/q^2\ll 1$ with the help of asymptotic
expansions~\cite{Smirnovbook}.  In the final expression the limit
  $m\to 0$ can be taken and the result coincides with the one obtain
  with the massless set-up. Possible infra-red singularities would
  manifest in $\ln m^2/q^2$ terms which are absent in our calculations.

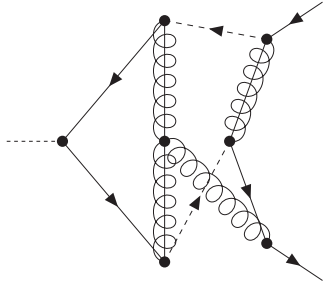
\begin{figure}
\begin{center}
\newcommand{\qvertex}[2]{%
\ifnum#1=-1 \ifnum#2=1 0 \else\ifnum#2=2 0\fi\fi\else
\ifnum#1=6
    \ifnum#2=1 75
\else\ifnum#2=2 10
\fi\fi\else
\ifnum#1=4
    \ifnum#2=1 75
\else\ifnum#2=2 140
\fi\fi\else
\ifnum#1=1
    \ifnum#2=1 130
\else\ifnum#2=2 130
\fi\fi\else
\ifnum#1=3
    \ifnum#2=1 130
\else\ifnum#2=2 20
\fi\fi\else
\ifnum#1=7
    \ifnum#2=1 75
\else\ifnum#2=2 75
\fi\fi\else
\ifnum#1=2
    \ifnum#2=1 20
\else\ifnum#2=2 75
\fi\fi\else
\ifnum#1=5
    \ifnum#2=1 110
\else\ifnum#2=2 75
\fi\fi\else
\fi\fi\fi\fi\fi\fi\fi\fi }
\SetScale{.7}
\begin{picture}(120,120)
\SetWidth{0.5}
\ArrowLine(\qvertex{4}{1},\qvertex{4}{2})(\qvertex{2}{1},\qvertex{2}{2})
\SetWidth{0.5}
\ArrowLine(\qvertex{2}{1},\qvertex{2}{2})(\qvertex{6}{1},\qvertex{6}{2})
\SetWidth{0.5}
\ArrowLine(\qvertex{5}{1},\qvertex{5}{2})(\qvertex{3}{1},\qvertex{3}{2})
\SetWidth{0.5}
\DashArrowLine(\qvertex{1}{1},\qvertex{1}{2})(\qvertex{4}{1},\qvertex{4}{2}){3}
\SetWidth{0.5}
\DashArrowLine(\qvertex{6}{1},\qvertex{6}{2})(\qvertex{5}{1},\qvertex{5}{2}){3}
\SetWidth{0.5}
\Line(\qvertex{1}{1},\qvertex{1}{2})(\qvertex{5}{1},\qvertex{5}{2})
\SetWidth{0.5}
\Gluon(\qvertex{1}{1},\qvertex{1}{2})(\qvertex{5}{1},\qvertex{5}{2}){6}{5}
\SetWidth{0.5}
\Line(\qvertex{7}{1},\qvertex{7}{2})(\qvertex{4}{1},\qvertex{4}{2})
\SetWidth{0.5}
\Gluon(\qvertex{7}{1},\qvertex{7}{2})(\qvertex{4}{1},\qvertex{4}{2}){6}{5}
\SetWidth{0.5}
\Line(\qvertex{6}{1},\qvertex{6}{2})(\qvertex{7}{1},\qvertex{7}{2})
\SetWidth{0.5}
\Gluon(\qvertex{6}{1},\qvertex{6}{2})(\qvertex{7}{1},\qvertex{7}{2}){6}{5}
\SetWidth{0.5}
\Gluon(\qvertex{7}{1},\qvertex{7}{2})(\qvertex{3}{1},\qvertex{3}{2}){6}{7.73702731645053}
\SetWidth{0.5}
\ArrowLine(160,150)(\qvertex{1}{1},\qvertex{1}{2})
\SetWidth{0.5}
\DashLine(-10,75)(\qvertex{2}{1},\qvertex{2}{2}){2}
\SetWidth{0.5}
\ArrowLine(\qvertex{3}{1},\qvertex{3}{2})(160,0)
\Vertex(\qvertex{6}{1},\qvertex{6}{2}){3}
\Vertex(\qvertex{4}{1},\qvertex{4}{2}){3}
\Vertex(\qvertex{1}{1},\qvertex{1}{2}){3}
\Vertex(\qvertex{3}{1},\qvertex{3}{2}){3}
\Vertex(\qvertex{7}{1},\qvertex{7}{2}){3}
\Vertex(\qvertex{2}{1},\qvertex{2}{2}){3}
\Vertex(\qvertex{5}{1},\qvertex{5}{2}){3}
\end{picture} 
\caption[]{\label{fig::qqep}Sample diagram for the three-loop $q\bar
  q\ep$ vertex where a non-vanishing trace with a single $\gamma_5$
  matrix occurs. The internal line styles are as in
  \fig{fig::majorana}.}
\end{center}
\end{figure}

Traces with a single $\gamma_5$ do not contribute to any of the
two-point functions that we calculated. They do contribute for some of
the three-point functions though, in particular the $q\tilde q\tilde g$, the
$\tilde g\tilde g\ep$, and the $q\bar q\ep$ vertex. An example
diagram for the latter vertex is shown in \fig{fig::qqep}. Such diagrams
contribute (among others) a color factor $d_R^{abcd}d_A^{abcd}$ (for
the notation, see 
Refs.\,\cite{vanRitbergen:1998pn,Jack:2007ni}), but they cancel against
the same factors from other sources in the final result for the
renormalization constants and the $\beta$ function.


The results for the renormalization constants can be obtained in
electronic form from Ref.~\cite{ttpdata}. It is straightforward
  to obtain  from them the three-loop \susy{}-\qcd{} $\beta$
function which reads
\begin{equation}
\begin{split}
& \beta(\alpha_s) =
-\sum_{n\geq 0} \left(\frac{\alpha_s}{\pi}\right)^{n+2}\beta_n\,,\\
   \beta_0 &=
         \frac{3}{4} C_A -\frac{ 1}{2}T_f\,,\\
   \beta_1 &=
        \frac{3}{8}\,C_A^2
       - T_f \left(  \frac{1}{2}\,C_F + \frac{1}{4} C_A \right)\,,\\
   \beta_2 &=
        \frac{21}{64} C_A^3 
       + T_f\left( \frac{1}{4}\,C_F^2 - \frac{13}{16}\,C_AC_F 
        - \frac{5}{16}\,C_A^2 \right)
       + T_f^2 \left( \frac{3}{8}\,C_F + \frac{1}{16}\,C_A \right)\,,
\end{split}
\label{eq::beta3l}
\end{equation}
where $C_F=(n_c^2-1)/(2n_c)$, $C_A=n_c$ are the Casimir operators of
SU($n_c$), and $2T_f=n_f$ is the number of quark flavors (which is
equal to the number of squark flavors).
The result in Eq.~(\ref{eq::beta3l}) is in full agreement with
Ref.~\cite{Jack:1996vg,Pickering:2001aq}.

The case $T_f=0$ (\susy{} Yang-Mills theory) has been treated before, both in
the superfield formalism~\cite{Jack:1996qq} and the diagrammatic
approach~\cite{Harlander:2006xq}. Full agreement has been found between
the two methods up to four-loop order.

The anomalous dimension for the quark mass is independent of any
\susy{}-breaking parameters and is given to three-loop  level
by
\begin{equation}
\begin{split}
& \gamma_{q}(\alpha_s) =
-\sum_{n\geq 0} \left(\frac{\alpha_s}{\pi}\right)^{n+1}\gamma^{q}_n\,,\\
   \gamma^{q}_0 & = \frac{1}{2} C_F\,,\\
\gamma^{q}_1 & = -\frac{1}{4} C_F^2 +\frac{3}{8} C_A C_F  - \frac{1}{4}
T_f C_F\,,\\
\gamma^{q}_2 & = \frac{1}{4} C_F^3 -\frac{3}{16} C_A C_F^2+\frac{3}{16}
C_A^2 C_F + T_f\left[\left(-\frac{1}{2}+\frac{3}{4}
  \zeta(3)\right)C_F^2  + \left(\frac{1}{16}-\frac{3}{4}\zeta(3)
  \right)C_A C_F\right] 
\\ & \quad -\frac{1}{8} T_f^2 C_F
\,,
\end{split}
\label{eq::gamma3l}
\end{equation}
where $\zeta$ is Riemann's zeta function with $\zeta(3)= 1.20206\ldots$.
The result of Eq.~(\ref{eq::gamma3l}) is in agreement with
  Ref.~\cite{Jack:1996qq}.

In Ref.~\cite{Jack:1997pa} it has been shown that the anomalous
  dimension of the gluino mass defined through
\begin{equation}
  \begin{split}
    \gamma_{{\tilde g}}(\alpha_s) &= -\sum_{n\geq 0}
    \left(\frac{\alpha_s}{\pi}\right)^{n+1}\gamma^{_{\tilde g}}_n\,,
  \end{split}
\end{equation}
can be related to the gauge $\beta$ function of
  Eq.~(\ref{eq::beta3l}) via 
\begin{equation}
\begin{split}
  \gamma^{_{\tilde g}}_n & =(n+1) \beta_{n}
  \,.
\end{split}
\end{equation}
We have  explicitly verified that  this relation holds through
three loops for the case of vanishing  \susy{}-breaking parameters except
the gluino mass.


\section{Conclusions}\label{sec::conclusions}

 The main result of this paper is a confirmation of the
consistency of \dred{} with \susy{} at three-loop order.

We have shown  that  the same result for the
 \susy{}-\qcd{}  $\beta$ function is obtained  
  from all  three-particle vertices, even the ones
involving \epscalar{}s which are used for the implementation of
\dred{}. Furthermore, we verify a predicted relation between the
$\beta$ function and the gluino mass anomalous dimension to three-loop order
and confirm the result for the quark mass anomalous dimension
present in the literature.

Along with this paper we provide the results for all two- and
three-point renormalization constants~\cite{ttpdata} which 
should be useful for other calculations within the \mssm{}. 

 
\paragraph{Acknowledgments.}
This work is supported by DFG through SFB/TR~9 and contract
   HA~2990/3-1, and by the Helmholtz Alliance ``Physics at the
    Terascale''. We would like to thank K.G.~Chetyrkin, 
    D.R.T.~Jones, P.~Kant, and D.~St\"ockinger for
  illuminating discussions and useful comments. 


\begin{appendix}

\section{Feynman Rules\label{app::frules}}


\newcommand{\qvertex}[2]{%
\ifnum#1=-1 \ifnum#2=1 0 \else\ifnum#2=2 0\fi\fi\else
\ifnum#1=6
    \ifnum#2=1 160
\else\ifnum#2=2 50
\fi\fi\else
\ifnum#1=1
    \ifnum#2=1 100
\else\ifnum#2=2 50
\fi\fi\else
\ifnum#1=3
    \ifnum#2=1 100
\else\ifnum#2=2 100
\fi\fi\else
\ifnum#1=4
    \ifnum#2=1 210
\else\ifnum#2=2 30
\fi\fi\else
\ifnum#1=2
    \ifnum#2=1 130
\else\ifnum#2=2 75
\fi\fi\else
\ifnum#1=5
    \ifnum#2=1 160
\else\ifnum#2=2 100
\fi\fi\else
\fi\fi\fi\fi\fi\fi\fi }


In this appendix we give the Feynman rules involving $\ep$-scalars and
flipped fermions. The notation is the same as in Appendix A.2 of
Ref.\,\cite{Harlander:2004tp}, except that all Lorentz indices occuring
there are understood as $d$-dimensional ones. Note that in that
paper, the sign of the $gg\tilde q\tilde q$ vertex should be
  reversed, and there is a typo for one of the $q\tilde q\tilde g$
vertices which should read\\[-2em]
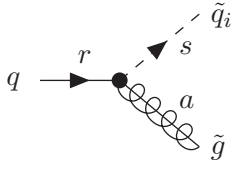
\begin{tabular}{ll}

%
%
\begin{picture}(100,70)(90,73)
\SetWidth{0.5}
\DashArrowLine(\qvertex{2}{1},\qvertex{2}{2})(\qvertex{5}{1},\qvertex{5}{2}){4}
\SetWidth{0.5}
\Line(\qvertex{6}{1},\qvertex{6}{2})(\qvertex{2}{1},\qvertex{2}{2})
\Gluon(\qvertex{6}{1},\qvertex{6}{2})(\qvertex{2}{1},\qvertex{2}{2}){4}{4}
\SetWidth{0.5}
\ArrowLine(100,75)(\qvertex{2}{1},\qvertex{2}{2})
\Vertex(\qvertex{2}{1},\qvertex{2}{2}){3}
\Text(165,50)[lc]{$\tilde g$}
\Text(165,100)[lc]{$\tilde q_i$}
\Text(93,75)[rc]{$q$}
\Text(153,90)[lt]{$s$}
\Text(153,65)[lb]{$a$}
\Text(120,82)[rb]{$r$}
\end{picture} &
$ig_sT^a_{sr}\sqrt{2}({\cal R}_{i1}^qP_L - {\cal R}_{i2}^qP_R)$\\


%
%
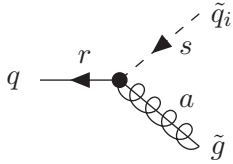
\begin{picture}(100,70)(90,73)
\SetWidth{0.5}
\DashArrowLine(\qvertex{5}{1},\qvertex{5}{2})(\qvertex{2}{1},\qvertex{2}{2}){4}
\SetWidth{0.5}
\Line(\qvertex{6}{1},\qvertex{6}{2})(\qvertex{2}{1},\qvertex{2}{2})
\Gluon(\qvertex{6}{1},\qvertex{6}{2})(\qvertex{2}{1},\qvertex{2}{2}){4}{4}
\SetWidth{0.5}
\ArrowLine(\qvertex{2}{1},\qvertex{2}{2})(100,75)
\Vertex(\qvertex{2}{1},\qvertex{2}{2}){3}
\Text(165,50)[lc]{$\tilde g$}
\Text(165,100)[lc]{$\tilde q_i$}
\Text(93,75)[rc]{$q$}
\Text(153,90)[lt]{$s$}
\Text(153,65)[lb]{$a$}
\Text(120,82)[rb]{$r$}
\end{picture} &
$ig_sT^a_{rs}\sqrt{2}({\cal R}_{i1}^qP_R - {\cal R}_{i2}^qP_L)$\\

\end{tabular}\\[3em]
Also, the four-gluon vertex did not contribute in
Ref.\,\cite{Harlander:2004tp}, so for completeness, let us quote it
here:\\[-2em]
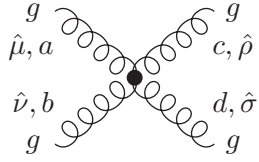
\begin{tabular}{ll}

%
%
\begin{picture}(100,70)(90,73)
\SetWidth{0.5}
\Gluon(\qvertex{2}{1},\qvertex{2}{2})(\qvertex{5}{1},\qvertex{5}{2}){4}{4}
\Gluon(\qvertex{6}{1},\qvertex{6}{2})(\qvertex{2}{1},\qvertex{2}{2}){4}{4}
\Gluon(\qvertex{3}{1},\qvertex{3}{2})(\qvertex{2}{1},\qvertex{2}{2}){4}{4}
\Gluon(\qvertex{2}{1},\qvertex{2}{2})(\qvertex{1}{1},\qvertex{1}{2}){4}{4}
\Vertex(\qvertex{2}{1},\qvertex{2}{2}){3}
\Text(165,50)[lc]{$g$}
\Text(165,100)[lc]{$g$}
\Text(95,100)[rc]{$g$}
\Text(95,50)[rc]{$g$}
\Text(160,92)[lt]{$c,\ddim{\rho}$}
\Text(160,60)[lb]{$d,\ddim{\sigma}$}
\Text(100,92)[rt]{$\ddim{\mu},a$}
\Text(100,60)[rb]{$\ddim{\nu},b$}
\end{picture} &
\begin{minipage}{12em}
\[
\begin{split}
-ig^2_sf^{abe}f^{cde}\left[ 
g^{\ddim{\mu}\ddim{\rho}}
g^{\ddim{\nu}\ddim{\sigma}}
-g^{\ddim{\mu}\ddim{\sigma}}
g^{\ddim{\nu}\ddim{\rho}}
  \right]
\\
-\,ig^2_sf^{ace}f^{bde}\left[
g^{\ddim{\mu}\ddim{\nu}}
g^{\ddim{\rho}\ddim{\sigma}}
-g^{\ddim{\mu}\ddim{\sigma}}
g^{\ddim{\nu}\ddim{\rho}}
  \right]
\\
-\,ig^2_sf^{ade}f^{bce}\left[
g^{\ddim{\mu}\ddim{\nu}}
g^{\ddim{\rho}\ddim{\sigma}}
-g^{\ddim{\mu}\ddim{\rho}}
g^{\ddim{\nu}\ddim{\sigma}}
  \right]
\end{split}
\]
\end{minipage}\\

\end{tabular}\\[3em]
The following are the tri-linear vertices involving \epscalar{}s
(all momenta incoming):\\[-2em]
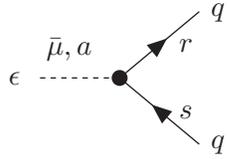
\begin{tabular}{ll}

%
%
\begin{picture}(100,70)(90,73)
\SetWidth{0.5}
\ArrowLine(\qvertex{2}{1},\qvertex{2}{2})(\qvertex{5}{1},\qvertex{5}{2})
\SetWidth{0.5}
\ArrowLine(\qvertex{6}{1},\qvertex{6}{2})(\qvertex{2}{1},\qvertex{2}{2})
\SetWidth{0.5}
\DashLine(100,75)(\qvertex{2}{1},\qvertex{2}{2}){2}
\Vertex(\qvertex{2}{1},\qvertex{2}{2}){3}
\Text(165,50)[lc]{$q$}
\Text(165,100)[lc]{$q$}
\Text(93,75)[rc]{$\ep$}
\Text(153,90)[lt]{$r$}
\Text(153,60)[lb]{$s$}
\Text(120,82)[rb]{$\edim\mu,a$}
\end{picture} &
$ig_sT^a_{rs}\gamma^{\edim\mu}$\\


%
%
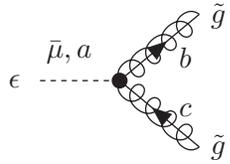
\begin{picture}(100,70)(90,73)
\SetWidth{0.5}
\ArrowLine(\qvertex{2}{1},\qvertex{2}{2})(\qvertex{5}{1},\qvertex{5}{2})
\Gluon(\qvertex{2}{1},\qvertex{2}{2})(\qvertex{5}{1},\qvertex{5}{2}){4}{4}
\SetWidth{0.5}
\ArrowLine(\qvertex{6}{1},\qvertex{6}{2})(\qvertex{2}{1},\qvertex{2}{2})
\Gluon(\qvertex{6}{1},\qvertex{6}{2})(\qvertex{2}{1},\qvertex{2}{2}){4}{4}
\SetWidth{0.5}
\DashLine(100,75)(\qvertex{2}{1},\qvertex{2}{2}){2}
\Vertex(\qvertex{2}{1},\qvertex{2}{2}){3}
\Text(165,50)[lc]{$\tilde g$}
\Text(165,100)[lc]{$\tilde g$}
\Text(93,75)[rc]{$\ep$}
\Text(153,88)[lt]{$b$}
\Text(153,62)[lb]{$c$}
\Text(120,82)[rb]{$\edim\mu,a$}
\end{picture} &
$g_sf^{abc}\gamma^{\edim\mu}$\\


%
%
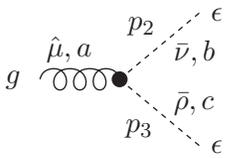
\begin{picture}(100,70)(90,73)
\SetWidth{0.5}
\DashLine(\qvertex{2}{1},\qvertex{2}{2})(\qvertex{5}{1},\qvertex{5}{2}){2}
\SetWidth{0.5}
\DashLine(\qvertex{6}{1},\qvertex{6}{2})(\qvertex{2}{1},\qvertex{2}{2}){2}
\SetWidth{0.5}
\Gluon(100,75)(\qvertex{2}{1},\qvertex{2}{2}){4}{3}
\Vertex(\qvertex{2}{1},\qvertex{2}{2}){3}
\Text(165,50)[lc]{$\ep$}
\Text(165,100)[lc]{$\ep$}
\Text(93,75)[rc]{$g$}
\Text(151,90)[lt]{$\edim\nu,b$}
\Text(151,62)[lb]{$\edim\rho,c$}
\Text(120,82)[rb]{$\ddim\mu,a$}
\Text(144,92)[rb]{$p_2$}
\Text(143,60)[rt]{$p_3$}
\end{picture} &
$g_sf^{abc}(p_2-p_3)^{\ddim\mu}g^{\edim\nu\edim\rho}$\\


%
%
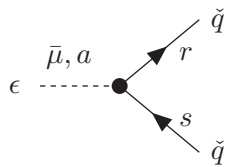
\begin{picture}(100,70)(90,73)
\SetWidth{0.5}
\ArrowLine(\qvertex{2}{1},\qvertex{2}{2})(\qvertex{5}{1},\qvertex{5}{2})
\SetWidth{0.5}
\ArrowLine(\qvertex{6}{1},\qvertex{6}{2})(\qvertex{2}{1},\qvertex{2}{2})
\SetWidth{0.5}
\DashLine(100,75)(\qvertex{2}{1},\qvertex{2}{2}){2}
\Vertex(\qvertex{2}{1},\qvertex{2}{2}){3}
\Text(165,50)[lc]{$\check q$}
\Text(165,100)[lc]{$\check q$}
\Text(93,75)[rc]{$\ep$}
\Text(153,90)[lt]{$r$}
\Text(153,60)[lb]{$s$}
\Text(120,82)[rb]{$\edim\mu,a$}
\end{picture} &
$-ig_sT^a_{sr}\gamma^{\edim\mu}$\\

\end{tabular}\\[3em]
As introduced in the main text, $\check q$ denotes a ``flipped'' quark
whose ``external flow'' (indicated by the arrow on the line) is opposite to
its charge flow. For clarity, the external flow has also been indicated
on the gluino lines in the $\ep\tilde g\tilde g$ vertex.
The other Feynman rules involving flipped quarks are
given as follows:\\[-2em]
\begin{tabular}{ll}

%
%
\begin{picture}(100,70)(90,73)
\SetWidth{0.5}
\ArrowLine(\qvertex{2}{1},\qvertex{2}{2})(\qvertex{5}{1},\qvertex{5}{2})
\SetWidth{0.5}
\ArrowLine(\qvertex{6}{1},\qvertex{6}{2})(\qvertex{2}{1},\qvertex{2}{2})
\SetWidth{0.5}
\Gluon(100,75)(\qvertex{2}{1},\qvertex{2}{2}){4}{3}
\Vertex(\qvertex{2}{1},\qvertex{2}{2}){3}
\Text(165,50)[lc]{$\check q$}
\Text(165,100)[lc]{$\check q$}
\Text(93,75)[rc]{$g$}
\Text(153,90)[lt]{$r$}
\Text(153,60)[lb]{$s$}
\Text(120,82)[rb]{$\ddim\mu,a$}
\end{picture} &
$-ig_sT^a_{sr}\gamma^{\ddim\mu}$\\


%
%
\begin{picture}(100,70)(90,73)
\SetWidth{0.5}
\DashArrowLine(\qvertex{2}{1},\qvertex{2}{2})(\qvertex{5}{1},\qvertex{5}{2}){4}
\SetWidth{0.5}
\Line(\qvertex{6}{1},\qvertex{6}{2})(\qvertex{2}{1},\qvertex{2}{2})
\Gluon(\qvertex{6}{1},\qvertex{6}{2})(\qvertex{2}{1},\qvertex{2}{2}){4}{4}
\SetWidth{0.5}
\ArrowLine(\qvertex{2}{1},\qvertex{2}{2})(100,75)
\Vertex(\qvertex{2}{1},\qvertex{2}{2}){3}
\Text(165,50)[lc]{$\tilde g$}
\Text(165,100)[lc]{$\tilde q_i$}
\Text(93,75)[rc]{$\check q$}
\Text(153,90)[lt]{$s$}
\Text(153,65)[lb]{$a$}
\Text(120,82)[rb]{$r$}
\end{picture} &
$ig_sT^a_{sr}\sqrt{2}({\cal R}_{i1}^qP_L - {\cal R}_{i2}^qP_R)$\\


%
%
\begin{picture}(100,70)(90,73)
\SetWidth{0.5}
\DashArrowLine(\qvertex{5}{1},\qvertex{5}{2})(\qvertex{2}{1},\qvertex{2}{2}){4}
\SetWidth{0.5}
\Line(\qvertex{6}{1},\qvertex{6}{2})(\qvertex{2}{1},\qvertex{2}{2})
\Gluon(\qvertex{6}{1},\qvertex{6}{2})(\qvertex{2}{1},\qvertex{2}{2}){4}{4}
\SetWidth{0.5}
\ArrowLine(100,75)(\qvertex{2}{1},\qvertex{2}{2})
\Vertex(\qvertex{2}{1},\qvertex{2}{2}){3}
\Text(165,50)[lc]{$\tilde g$}
\Text(165,100)[lc]{$\tilde q_i$}
\Text(93,75)[rc]{$\check q$}
\Text(153,90)[lt]{$s$}
\Text(153,65)[lb]{$a$}
\Text(120,82)[rb]{$r$}
\end{picture} &
$ig_sT^a_{rs}\sqrt{2}({\cal R}_{i1}^qP_R - {\cal R}_{i2}^qP_L)$\\

\end{tabular}\\[3em]
The quartic vertices involving \epscalar{}s are\\[-2em]
\begin{tabular}{ll}

%
%
\begin{picture}(100,70)(90,73)
\SetWidth{0.5}
\DashLine(\qvertex{2}{1},\qvertex{2}{2})(\qvertex{5}{1},\qvertex{5}{2}){2}
\SetWidth{0.5}
\DashLine(\qvertex{6}{1},\qvertex{6}{2})(\qvertex{2}{1},\qvertex{2}{2}){2}
\SetWidth{0.5}
\Gluon(\qvertex{3}{1},\qvertex{3}{2})(\qvertex{2}{1},\qvertex{2}{2}){4}{4}
\Gluon(\qvertex{2}{1},\qvertex{2}{2})(\qvertex{1}{1},\qvertex{1}{2}){4}{4}
\Vertex(\qvertex{2}{1},\qvertex{2}{2}){3}
\Text(165,50)[lc]{$\ep$}
\Text(165,100)[lc]{$\ep$}
\Text(95,100)[rc]{$g$}
\Text(95,50)[rc]{$g$}
\Text(160,92)[lt]{$c,\edim{\rho}$}
\Text(160,60)[lb]{$d,\edim{\sigma}$}
\Text(100,92)[rt]{$\ddim{\mu},a$}
\Text(100,60)[rb]{$\ddim{\nu},b$}
\end{picture} &
\begin{minipage}{12em}
\[
\begin{split}
-\,ig^2_s\,
g^{\ddim{\mu}\ddim{\nu}}
g^{\edim{\rho}\edim{\sigma}}
\left[
f^{ace}f^{bde} + f^{ade}f^{bce}
  \right]
\end{split}
\]
\end{minipage}\\


%
%
\begin{picture}(100,70)(90,73)
\SetWidth{0.5}
\DashArrowLine(\qvertex{2}{1},\qvertex{2}{2})(\qvertex{5}{1},\qvertex{5}{2}){4}
\DashArrowLine(\qvertex{6}{1},\qvertex{6}{2})(\qvertex{2}{1},\qvertex{2}{2}){4}
\DashLine(\qvertex{3}{1},\qvertex{3}{2})(\qvertex{2}{1},\qvertex{2}{2}){2}
\DashLine(\qvertex{2}{1},\qvertex{2}{2})(\qvertex{1}{1},\qvertex{1}{2}){2}
\Vertex(\qvertex{2}{1},\qvertex{2}{2}){3}
\Text(165,50)[lc]{$\tilde q_j$}
\Text(165,100)[lc]{$\tilde q_i$}
\Text(95,100)[rc]{$\ep$}
\Text(95,50)[rc]{$\ep$}
\Text(153,90)[lt]{$r$}
\Text(153,60)[lb]{$s$}
\Text(108,90)[rt]{$\edim\nu,b$}
\Text(108,60)[rb]{$\edim\mu,a$}
\end{picture} &
$
ig_s^2\left\{T^a,T^b\right\}_{rs}g^{\edim\mu\edim\nu}\delta_{ij} =
ig_s^2\left(\frac{1}{3}\delta_{ab}\delta_{rs} +
d_{abc}T_{rs}^c\right)g^{\edim\mu\edim\nu}\delta_{ij}
$\\


%
%
\begin{picture}(100,70)(90,73)
\SetWidth{0.5}
\DashLine(\qvertex{2}{1},\qvertex{2}{2})(\qvertex{5}{1},\qvertex{5}{2}){2}
\SetWidth{0.5}
\DashLine(\qvertex{6}{1},\qvertex{6}{2})(\qvertex{2}{1},\qvertex{2}{2}){2}
\SetWidth{0.5}
\DashLine(\qvertex{3}{1},\qvertex{3}{2})(\qvertex{2}{1},\qvertex{2}{2}){2}
\DashLine(\qvertex{2}{1},\qvertex{2}{2})(\qvertex{1}{1},\qvertex{1}{2}){2}
\Vertex(\qvertex{2}{1},\qvertex{2}{2}){3}
\Text(165,50)[lc]{$\ep$}
\Text(165,100)[lc]{$\ep$}
\Text(95,100)[rc]{$\ep$}
\Text(95,50)[rc]{$\ep$}
\Text(160,92)[lt]{$c,\edim{\rho}$}
\Text(160,60)[lb]{$d,\edim{\sigma}$}
\Text(100,92)[rt]{$\edim{\mu},a$}
\Text(100,60)[rb]{$\edim{\nu},b$}
\end{picture} &
\begin{minipage}{12em}
\[
\begin{split}
-ig^2_sf^{abe}f^{cde}\left[ 
g^{\edim{\mu}\edim{\rho}}
g^{\edim{\nu}\edim{\sigma}}
-g^{\edim{\mu}\edim{\sigma}}
g^{\edim{\nu}\edim{\rho}}
  \right]
\\
-\,ig^2_sf^{ace}f^{bde}\left[
g^{\edim{\mu}\edim{\nu}}
g^{\edim{\rho}\edim{\sigma}}
-g^{\edim{\mu}\edim{\sigma}}
g^{\edim{\nu}\edim{\rho}}
  \right]
\\
-\,ig^2_sf^{ade}f^{bce}\left[
g^{\edim{\mu}\edim{\nu}}
g^{\edim{\rho}\edim{\sigma}}
-g^{\edim{\mu}\edim{\rho}}
g^{\edim{\nu}\edim{\sigma}}
  \right]
\end{split}
\]
\end{minipage}\\

\end{tabular}\\[3em]
\end{appendix}


\def\app#1#2#3{{\it Act.~Phys.~Pol.~}\jref{\bf B #1}{#2}{#3}}
\def\apa#1#2#3{{\it Act.~Phys.~Austr.~}\jref{\bf#1}{#2}{#3}}
\def\annphys#1#2#3{{\it Ann.~Phys.~}\jref{\bf #1}{#2}{#3}}
\def\cmp#1#2#3{{\it Comm.~Math.~Phys.~}\jref{\bf #1}{#2}{#3}}
\def\cpc#1#2#3{{\it Comp.~Phys.~Commun.~}\jref{\bf #1}{#2}{#3}}
\def\epjc#1#2#3{{\it Eur.\ Phys.\ J.\ }\jref{\bf C #1}{#2}{#3}}
\def\fortp#1#2#3{{\it Fortschr.~Phys.~}\jref{\bf#1}{#2}{#3}}
\def\ijmpc#1#2#3{{\it Int.~J.~Mod.~Phys.~}\jref{\bf C #1}{#2}{#3}}
\def\ijmpa#1#2#3{{\it Int.~J.~Mod.~Phys.~}\jref{\bf A #1}{#2}{#3}}
\def\jcp#1#2#3{{\it J.~Comp.~Phys.~}\jref{\bf #1}{#2}{#3}}
\def\jetp#1#2#3{{\it JETP~Lett.~}\jref{\bf #1}{#2}{#3}}
\def\jphysg#1#2#3{{\small\it J.~Phys.~G~}\jref{\bf #1}{#2}{#3}}
\def\jhep#1#2#3{{\small\it JHEP~}\jref{\bf #1}{#2}{#3}}
\def\mpl#1#2#3{{\it Mod.~Phys.~Lett.~}\jref{\bf A #1}{#2}{#3}}
\def\nima#1#2#3{{\it Nucl.~Inst.~Meth.~}\jref{\bf A #1}{#2}{#3}}
\def\npb#1#2#3{{\it Nucl.~Phys.~}\jref{\bf B #1}{#2}{#3}}
\def\nca#1#2#3{{\it Nuovo~Cim.~}\jref{\bf #1A}{#2}{#3}}
\def\plb#1#2#3{{\it Phys.~Lett.~}\jref{\bf B #1}{#2}{#3}}
\def\prc#1#2#3{{\it Phys.~Reports }\jref{\bf #1}{#2}{#3}}
\def\prd#1#2#3{{\it Phys.~Rev.~}\jref{\bf D #1}{#2}{#3}}
\def\pR#1#2#3{{\it Phys.~Rev.~}\jref{\bf #1}{#2}{#3}}
\def\prl#1#2#3{{\it Phys.~Rev.~Lett.~}\jref{\bf #1}{#2}{#3}}
\def\pr#1#2#3{{\it Phys.~Reports }\jref{\bf #1}{#2}{#3}}
\def\ptp#1#2#3{{\it Prog.~Theor.~Phys.~}\jref{\bf #1}{#2}{#3}}
\def\ppnp#1#2#3{{\it Prog.~Part.~Nucl.~Phys.~}\jref{\bf #1}{#2}{#3}}
\def\rmp#1#2#3{{\it Rev.~Mod.~Phys.~}\jref{\bf #1}{#2}{#3}}
\def\sovnp#1#2#3{{\it Sov.~J.~Nucl.~Phys.~}\jref{\bf #1}{#2}{#3}}
\def\sovus#1#2#3{{\it Sov.~Phys.~Usp.~}\jref{\bf #1}{#2}{#3}}
\def\tmf#1#2#3{{\it Teor.~Mat.~Fiz.~}\jref{\bf #1}{#2}{#3}}
\def\tmp#1#2#3{{\it Theor.~Math.~Phys.~}\jref{\bf #1}{#2}{#3}}
\def\yadfiz#1#2#3{{\it Yad.~Fiz.~}\jref{\bf #1}{#2}{#3}}
\def\zpc#1#2#3{{\it Z.~Phys.~}\jref{\bf C #1}{#2}{#3}}
\def\ibid#1#2#3{{ibid.~}\jref{\bf #1}{#2}{#3}}
\def\otherjournal#1#2#3#4{{\it #1}\jref{\bf #2}{#3}{#4}}

\newcommand{\jref}[3]{{\bf #1}, #3 (#2)}
\newcommand{\bibentry}[4]{#1, {\it #2}, #3\ifthenelse{\equal{#4}{}}{}{, }#4.}
\newcommand{\hepph}[1]{[hep-ph/#1]}
\newcommand{\mathph}[1]{[math-ph/#1]}
\newcommand{\arxiv}[1]{{\tt arXiv:#1}}



\begin{thebibliography}{99}

%
%

\bibitem{Jack:1996vg}
\bibentry{I.~Jack, D.R.T.~Jones, C.G.~North}
{$N = 1$ supersymmetry and the three loop gauge beta function}
{\plb{386}{1996}{138}}
{\hepph{9606323}}

\bibitem{Pickering:2001aq}
\bibentry{A.G.M.~Pickering, J.A.~Gracey, D.R.T.~Jones}
{Three loop gauge beta-function for the most general single  gauge-coupling
theory}
{\plb{510}{2001}{347}}
{\hepph{0104247}}

\bibitem{Susskind:1978ms}
\bibentry{L.~Susskind}
  {Dynamics Of Spontaneous Symmetry Breaking In The Weinberg-Salam Theory}
{\prd{20}{1979}{2619}}
{}

\bibitem{Weinberg:1979bn}
\bibentry{  S.~Weinberg}
  {Implications Of Dynamical Symmetry Breaking: An Addendum}
  {\prd{19}{1979} {1277}}
{}

\bibitem{tHooft:1979bh}
\bibentry{G.~'t Hooft}
{Naturalness, Chiral Symmetry, And Spontaneous Chiral Symmetry Breaking}
{{\it NATO Adv.\ Study Inst.\ Ser.\ B Phys.}  {\bf 59} (1980) 135}
{}

\bibitem{Ellis:1990wk}
\bibentry{J.R.~Ellis, S.~Kelley, D.V.~Nanopoulos}
{Probing the desert using gauge coupling unification}
{\plb{260}{1991}{131}}
{}

\bibitem{Amaldi:1991cn}
\bibentry{U.~Amaldi, W.~de Boer, H.~F\"urstenau}
{Comparison of grand unified theories with electroweak and strong coupling
constants measured at LEP}
{\plb{260}{1991}{447}}
{}


\bibitem{Langacker:1991an}
\bibentry{P.~Langacker, M.~x.~Luo}
{Implications of precision electroweak experiments for $M_t$, $\rho_{0}$,
$\sin^2\theta_W$ and grand unification}
{\prd{44}{1991}{817}}
{}

\bibitem{Nilles:1983ge}
\bibentry{H.P.~Nilles}
{Supersymmetry, Supergravity And Particle Physics}
{\pr{110}{1984}{1}}
{}

\bibitem{Harlander:2007wh}
\bibentry{R.V.~Harlander, L.~Mihaila, M.~Steinhauser}
{Running of $\alpha_s$ and $m_b$ in the MSSM}
{\prd{76}{2007}{055002}}
{\arxiv{0706.2953 [hep-ph]}}

\bibitem{Siegel:1979wq}
\bibentry{W.~Siegel}
{Supersymmetric Dimensional Regularization Via Dimensional Reduction}
{\plb{84}{1979}{193}}
{}

\bibitem{Siegel:1980qs}
\bibentry{W.~Siegel}
{Inconsistency Of Supersymmetric Dimensional Regularization}
{\plb{94}{1980}{37}}
{}

\bibitem{Aguilar-Saavedra:2005pw}
\bibentry{J.A.~Aguilar-Saavedra {\it et al.}}
{Supersymmetry parameter analysis: SPA convention and project}
{\epjc{46}{2006}{43}}
{\hepph{0511344}}

\bibitem{Stockinger:2005gx}
\bibentry{D.~St\"ockinger}
{Regularization by dimensional reduction: Consistency, quantum action
principle, and supersymmetry}
{\jhep{0503}{2005}{076}}
{\hepph{0503129}}

\bibitem{vanRitbergen:1997va}
\bibentry{T.~van Ritbergen, J.A.M.~Vermaseren, S.A.~Larin}
{The four-loop beta function in quantum chromodynamics}
{\plb{400}{1997}{379}}
{\hepph{9701390}}

\bibitem{Chetyrkin:1997dh}
\bibentry{K.G.~Chetyrkin}
{Quark mass anomalous dimension to $\order{\alpha_s^4}$}
{\plb{404}{1997}{161}}
{\hepph{9703278}}

\bibitem{Vermaseren:1997fq}
  \bibentry{J.A.~Vermaseren, S.A.~Larin, T.~van Ritbergen}
{The 4-loop quark mass anomalous dimension and the invariant quark  mass}
{\plb{405}{1997}{327}}
{\hepph{9703284}}

\bibitem{Czakon:2004bu}
\bibentry{M.~Czakon}
{The four-loop QCD beta-function and anomalous dimensions}
{\npb{710}{2005}{485}}
{\hepph{0411261}}

\bibitem{Harlander:2006xq}
\bibentry{R.V.~Harlander, D.R.T.~Jones, P.~Kant, L.~Mihaila, M.~Steinhauser}
{Four-loop beta function and mass anomalous dimension in Dimensional
Reduction}
{\jhep{0612}{2006}{024}}
{\hepph{0610206}}

\bibitem{Novikov:1983uc}
\bibentry{V.A.~Novikov, M.A.~Shifman, A.I.~Vainshtein, V.I.~Zakharov}
{Exact Gell-Mann-Low Function Of Supersymmetric Yang-Mills Theories From
Instanton Calculus}
{\npb{229}{1983}{381}}
{}

\bibitem{Hisano:1997ua}
\bibentry{J.~Hisano and M.A.~Shifman}
{Exact results for soft supersymmetry breaking parameters in  supersymmetric
gauge theories}
{\prd{56}{1997}{5475}}
{\hepph{9705417}}

\bibitem{Jack:1997pa}
\bibentry{I.~Jack and D.R.T.~Jones}
{The gaugino beta-function}
{\plb{415}{1997}{383}}
{\hepph{9709364}}

\bibitem{Jack:1996qq}
\bibentry{ I.~Jack, D.~R.~T.~Jones and C.~G.~North}
  {$N=1$ supersymmetry and the three loop anomalous dimension for the chiral
  superfield}
{\npb{473}{1996}{308}}
{\hepph{9603386}} 

\bibitem{Avdeev:1981vf}
\bibentry{  L.~V.~Avdeev, G.~A.~Chochia and A.~A.~Vladimirov}
  {On The Scope Of Supersymmetric Dimensional Regularization}
{\plb{ 105}{1981}{272}}
{}

\bibitem{Harlander:2004tp}
\bibentry{R.V.~Harlander and M.~Steinhauser}
{Supersymmetric Higgs production in gluon fusion at next-to-leading order}
{\jhep{0409}{2004}{066}}
{\hepph{0409010}}

\bibitem{Harlander:2006rj}
\bibentry{R.~Harlander, P.~Kant, L.~Mihaila, M.~Steinhauser}
{Dimensional reduction applied to QCD at three loops}
{\jhep{09}{2006}{053}}
{\hepph{0607240}}

\bibitem{Jack:2007ni}
\bibentry{I.~Jack, D.R.T.~Jones, P.~Kant, L.~Mihaila}
{The four-loop DRED gauge beta-function and fermion mass anomalous dimension
for general gauge groups}
{\jhep{0709}{2007}{058}}
{\arxiv{0707.3055 [hep-th]}}

\bibitem{Denner:1992me}
\bibentry{A.~Denner, H.~Eck, O.~Hahn, J.~K\"ublbeck}
{Compact Feynman rules for Majorana fermions}
{\plb{291}{1992}{278}}
{}

\bibitem{Denner:1992vza}
\bibentry{A.~Denner, H.~Eck, O.~Hahn, J.~K\"ublbeck}
{Feynman rules for fermion number violating interactions}
{\npb{387}{1992}{467}}
{}

\bibitem{Nogueira:1993ex}
\bibentry{P.~Nogueira}
{Automatic Feynman graph generation}
{\jcp{105}{1993}{279}}
{}

\bibitem{Harlander:2008ju}
\bibentry{R.V.~Harlander, P.~Kant, L.~Mihaila, M.~Steinhauser}
{Higgs boson mass in supersymmetry to three loops}
{\prl{100}{2008}{191602}; (E) \ibid{101}{2008}{039901}}
{\arxiv{0803.0672 [hep-ph]}}

\bibitem{Larin:1991fz}
\bibentry{S.A.~Larin, F.V.~Tkachov, J.A.~Vermaseren}
{The FORM version of MINCER}
{NIKHEF-H-91-18}
{}

\bibitem{exp} 
%
\bibentry{T.~Seidensticker}
{Automatic application of successive asymptotic expansions of Feynman
diagrams}
{\hepph{9905298}}
{}\\
%
\bibentry{R.~Harlander, T.~Seidensticker, M.~Steinhauser}
{Corrections of ${\cal O}(\alpha \alpha_s)$ to the decay of the $Z$ boson  
into bottom quarks}
{\plb{426}{1998}{125}}
{\hepph{9712228}}

\bibitem{Vermaseren:2000nd}
\bibentry{J.A.~Vermaseren}
{New features of {\abbrev FORM}}
{\mathph{0010025}}
{}

\bibitem{vanRitbergen:1998pn}
\bibentry{T.~van Ritbergen, A.N.~Schellekens, J.A.M.~Vermaseren}
{Group theory factors for Feynman diagrams}
{\ijmpa{14}{41}{1999}}
{\hepph{9802376}}


\bibitem{Steinhauser:1998cm}
\bibentry{M.~Steinhauser}
{Higgs decay into gluons up to $\order{\alpha_s^3 G_F m_t^2}$}
{\prd{59}{1999}{054005}}
{\hepph{9809507}}


\bibitem{Steinhauser:2002rq}
\bibentry{M.~Steinhauser}
{Results and techniques of multi-loop calculations}
{\pr{364}{2002}{247}}
{\hepph{0201075}}

\bibitem{Bauer:2008bj}
\bibentry{A.~Bauer, L.~Mihaila, J.~Salomon}
{Matching coefficients for $\alpha_s$ and $m_b$ to ${\cal
  O}(\alpha_s^2)$ in the MSSM}
{\jhep{0902}{2009}{037}}
{\arxiv{0810.5101 [hep-ph]}}

\bibitem{Smirnovbook}
\bibentry{V.A.~Smirnov}
{Applied asymptotic expansions in momenta and masses}
{Springer Tracts in Modern Physics, Vol.~177 (2002), ISBN 3-540-42334-6}
{}

\bibitem{ttpdata}
{\tt http://www-ttp.particle.uni-karlsruhe.de/Progdata/ttp09/ttp09-13.}


\end{thebibliography}
\end{document}